\documentstyle[ltwol]{article}

\input{psfig}

\def\NPB{{\em Nucl. Phys.} B}
\def\PLB{{\em Phys. Lett.}  B}
\def\PRL{\em Phys. Rev. Lett.}
\def\PR{\em Phys. Rep.}
\def\PRD{{\em Phys. Rev.} D}
\def\ZPC{{\em Z. Phys.} C}

\def\nn{\noindent}
\def\ie{{\it i.e.}}
\def\eg{{\it e.g.}}

\def\etal{{\it et al.}}

\def\be{\begin{equation}}
\def\ee{\end{equation}}
\def\bea{\begin{eqnarray}}
\def\eea{\end{eqnarray}}

\begin{document}

\onecolumn

\vbox{ \large
\begin{flushright}
SLAC-PUB-7986\\
February 1999\\
\end{flushright}

\vspace{1in}

\begin{center}
{Top Quark Production at the Tevatron: Probing Anomalous Chromomagnetic 
Moments and Theories of Low Scale Gravity}
\medskip
\vskip .3cm

{\large Thomas G. Rizzo }\\
Stanford Linear Accelerator Center, Stanford, CA 94309, USA\\
\vskip 1.0cm

\vskip3cm

{\bf Abstract}\\
\end{center}
We review the effects of an anomalous chromomagnetic moment on top quark pair 
production at the Tevatron for Run II. Apart from an overall rescaling by 
the total cross section these couplings have little influence on top 
production characteristics. We then examine how the exchange of a 
Kaluza-Klein tower of gravitons arising from the low scale quantum gravity 
scenario of Arkani-Hamed, Dimopoulos and Dvali influences top pair cross 
sections and several associated kinematical distributions. The data that will 
be available with Run II luminosities is shown to be sensitive to values of 
the effective Planck scale of order 1 TeV or more. 

\vspace*{2.0in}
\noindent 
Extended version of a talk given at {\it Physics at Run II: Workshop on Top 
Physics}, Fermilab, 16-18 October, 1998
\vskip 1.7cm

\noindent $^*$Work supported by the U.S. Department of Energy under contract
DE-AC03-76SF00515. 

\thispagestyle{empty}
}
\newpage


\title{{TOP QUARK PRODUCTION AT THE TEVATRON: PROBING ANOMALOUS CHROMOMAGNETIC 
MOMENTS AND THEORIES OF LOW SCALE GRAVITY}}

\author{ {THOMAS G. RIZZO}}

\address{Stanford Linear Accelerator Center, Stanford University, Stanford, 
CA 94309, USA\\E-mail: rizzo@slac.stanford.edu}

\twocolumn[\maketitle\abstracts{
We review the effects of an anomalous chromomagnetic moment on top quark pair 
production at the Tevatron for Run II. Apart from an overall rescaling by 
the total cross section these couplings have little influence on top 
production characteristics. We then examine how the exchange of a 
Kaluza-Klein tower of gravitons arising from the low scale quantum gravity 
scenario of Arkani-Hamed, Dimopoulos and Dvali influences top pair cross 
sections and several associated kinematical distributions. The data that will 
be available with Run II luminosities is shown to be sensitive to values of 
the effective Planck scale of order 1 TeV or more. 
}]

\section{Introduction}

The production of top quarks at the Tevatron and other high energy colliders 
may be sensitive to new physics beyond the electroweak scale. This new 
physics can take many forms but in a very large 
class of models it can be parameterized 
by a set of non-renormalizable operators. In this paper we consider how the 
existence of two of these operators, one of dimension-six while the other is 
dimension-eight, modifies the production properties of the top at the 
Tevatron. The first operator induces an anomalous chromomagnetic dipole moment 
coupling between the top and gluons and has already been studied in some 
detail. The second operator, which has yet to be examined, occurs in a new 
theory of 
low energy quantum gravity~{\cite {nima}} and has nothing to do with the 
strong interactions but arises from the exchange of a Kaluza-Klein tower of 
gravitons. While the underlying physics behind these operators is completely 
different their influence on top production can be 
analyzed in a similar fashion. However, the two scenarios lead to remarkably 
different predictions which can be tested at Run II.

\section{Dimension-Six Operators and Chromomagnetic Moments}

Let us first consider the set of dimension-six operators that can 
lead to a modification of the coupling of a gluon to $t\bar t$. Demanding 
$P$, $C$, and $CP$ conservation, one can show~{\cite {bing}} that there are 
only two Standard Model(SM), \ie, $SU(3)_c\times SU(2)_L\times U(1)_Y$, 
gauge invariant operators of dimension-six which can be constructed out of SM 
fields which lead to new physics at the corresponding 
$t \bar t g$ vertex. (It will be assumed that the lighter quarks do not 
experience these new interactions.) These two operators can be written as
\begin{eqnarray}
{\cal L}_{int}&=& {c_1\over{\Lambda^2}}[\bar q_L\gamma^\mu T^a D^\nu q_L+
\bar t_R\gamma^\mu T^a D^\nu t_R+h.c.]G^a_{\mu\nu}\nonumber \\
&+& {c_2\over{\Lambda^2}}[\bar q_L\sigma^{\mu\nu}T^a\tilde \phi
+h.c.]G^a_{\mu\nu}\,,
\end{eqnarray}
where $\Lambda$ is the scale of the new physics, the $c_i$ are 
unknown coefficients 
of order unity, $q_L$ is the left-handed doublet containing the top and bottom 
quarks, $T^a$ are the usual $SU(3)_c$ 
generators, $D^\nu$ is the SM covariant derivative, $G^a_{\mu\nu}$ is the 
gluon field strength tensor, and $\tilde \phi=i\tau_2\phi^*$ is the conjugate 
of the SM Higgs field. (If one gives up parity conservation or charge 
conjugation invariance only one additional 
new operator is introduced, whereas, if we surrender all of $P$, $C$ and 
$CP$ four additional operators arise~{\cite {bing}}.) 
Once the Higgs field obtains a vev, $v/\sqrt 2$, and we place the 
top quark pair on shell the resulting effective $t\bar t g$ vertex, including 
the usual piece from QCD, can be written very simply as

\begin{equation}
{\cal L}_{eff}=g_s\bar t T_a \left(F_v \gamma_\mu G_a^\mu-{\kappa \over
{4m_t}}\sigma_{\mu\nu}G_a^{\mu\nu}\right)t  \,,
\end{equation}
where $g_s$ is the usual $SU(3)_c$ coupling, $G_a^\mu$ is the gluon field 
and we make the identifications $\kappa=2\sqrt 2m_tvc_2/\Lambda^2$ and 
$F_v=1-c_1q^2/\Lambda^2$ with $q^2$ being the invariant mass of the $t\bar t$ 
pair. The parameter $\kappa$ can be identified as the anomalous 
chromomagnetic moment of the top quark; we observe that when $\kappa \neq 0$ 
a direct $ggt\bar t$ four-point function is induced as a result of 
gauge invariance. Note that for $\Lambda$=1 TeV and $c_2$ of order unity 
values of $\kappa$ of order 0.1 are possible~{\cite {est}}.

\vspace*{-0.5cm}
\nn
\begin{figure}[htbp]
\centerline{
\psfig{figure=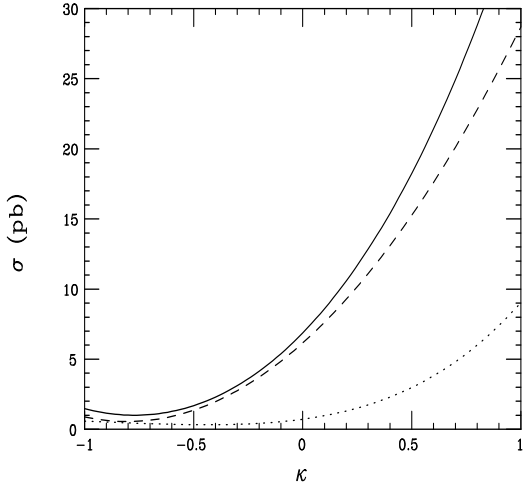,height=8.0cm,width=8.5cm,angle=-90}}
\vspace*{-0.6cm}
\caption{Cross section for $t\bar t$ production (solid) at Tevatron for Run II 
as a function of $\kappa$. The part of the cross section arising from 
$gg(q\bar q)$ annihilation is shown by the dotted(dashed) curve.}
\label{fig1}
\end{figure}
\vspace*{0.1mm}

The effects of $F_v\neq 1$ on top quark pair production via 
$q\bar q \to t\bar t$ are relatively straightforward to analyze~{\cite {bing}} 
since they simply scale the strength of the ordinary QCD coupling by an 
additional $q^2$-dependent amount. $F_v$ is thus seen to act like a simple 
form factor. The effects 
associated with $\kappa$ were examined much earlier~{\cite {old}} as the 
existence of $\kappa \neq 0$ was originally motivated by problems in flavor 
physics~{\cite {alex}} as well as early high top cross section measurements. 
As one might guess, and as was explicitly demonstrated~{\cite {old}} by 
Hikasa \etal, the effects of these two operators will 
be easily distinguishable at the Tevatron. 

In order to examine the effects due to $\kappa \neq 0$, we must calculate the 
parton-level $q\bar q \to t\bar t$ (which dominates at Tevatron energies) and 
$gg \to t\bar t$ differential cross 
sections. For the $q\bar q$ case one obtains~{\cite {veryold}}
\begin{eqnarray}
{d\sigma_{q\bar q}\over {d\hat t}}&=&{8\pi\alpha_s^2\over {27\hat
s^2}}\left[
\left(1+{2m_t^2\over {\hat s}}\right)+3\kappa+\kappa^2\left({\hat s\over
{8m_t^2}}\right)\right. \nonumber \\
&+&\left.{\beta^2 \over {4}}(3z^2-1)\left(1-{\hat s\over
{4m_t^2}}\kappa^2\right)\right]\,,
\end{eqnarray}
with $\hat s$ being the parton level center of mass energy, 
$\beta^2=1-4m_t^2/\hat s$, and $z$ being the cosine of the corresponding 
scattering angle, $\theta^*$. $m_t$ will be taken to be 175 GeV in the 
numerical discussion below. The corresponding cross section for 
$gg \to t\bar t$ is somewhat lengthy~{\cite {old}} and not very enlightening 
except for the fact that it is quartic in $\kappa$.
Both of these cross section expressions have also been 
generalized~{\cite {old}} to allow for a $CP$-violating anomalous 
chromoelectric moment, $\tilde \kappa$ which can appear in cross section 
expressions in even powers only since cross sections are not $CP$-violating 
observables. Such a coupling arises from the 
$CP$-violating analogue of the operator proportional to $c_2$ in Eq.1 when the 
gluon field strength is replaced by its dual. 

Several comments about the above cross section expression, which also 
apply to the $gg$ initiated case, are now relevant. First, we 
see that $\kappa \neq 0$ modifies both 
the total cross section as well as the angular distribution. Second, the 
influence of $\kappa$ grows rapidly with increasing $\hat s/m_t^2$ which 
implies that the effects for a fixed value of $\kappa$ are more likely to be 
observed as the collider center of mass energy increases. Thus we may expect 
that $\kappa$ is more easily probed at the LHC than it is at the Tevatron. 
Third, the 
differential cross section is sensitive to the {\it sign} of $\kappa$ 
through the interference with the SM coupling and this can lead to both a 
significant suppression or enhancement in the production rate. 

\vspace*{-0.5cm}
\nn
\begin{figure}[htbp]
\centerline{
\psfig{figure=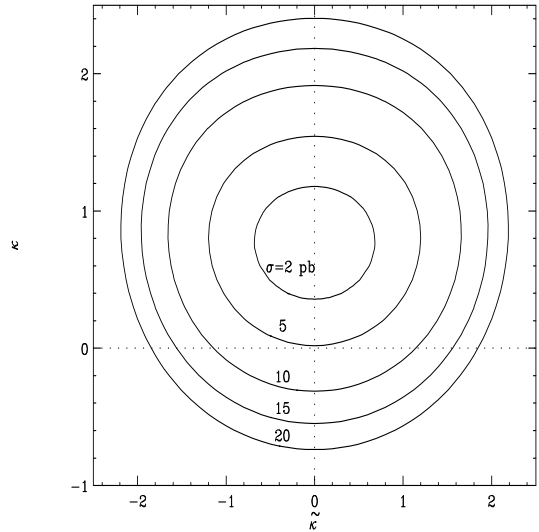,height=7cm,width=7cm,angle=-0}}
\vspace*{-0.0cm}
\caption{Same as Fig. 1 but now as a function of both $\kappa$ and 
$\tilde \kappa$ from the work of Cheung in Ref. 4. The sign of $\kappa$ is 
here opposite to that in Fig. 1.}
\label{fig1k}
\end{figure}
\vspace*{0.1mm}

To calculate the cross sections and associated distributions we employ 
the tree-level expression above, and the corresponding one for 
$gg\to t\bar t$, which are then corrected by K-factors~{\cite {old}} 
in order to include the effects of higher order QCD corrections. For 
definiteness we use the CTEQ4M parton distribution functions~{\cite {cteq}} 
although our numerical results will not be sensitive to this particular 
choice. The results of this cross section calculation for the Tevatron 
with $\sqrt s=2$ TeV is shown in Fig.1. We see immediately that if 
$\kappa$ is $\neq 0$ and of appreciable magnitude the top cross section can be 
significantly higher or lower than in the SM depending upon the sign of 
$\kappa$. This result can be used to constrain the value of $\kappa$ provided 
no deviation from the SM prediction is observed. For example, assuming the SM 
cross section prediction was obtained experimentally and that the combined 
theoretical and experimental uncertainty on this measurement was at the 
level of $20\%$, we would deduce the bound $-0.10 \leq \kappa \leq 0.12$, 
which is beginning to probe an interesting range. 
Unfortunately, as we will soon see, it will be difficult for the Tevatron to do 
much better than this. 

\vspace*{-0.5cm}
\nn
\begin{figure}[htbp]
\centerline{
\psfig{figure=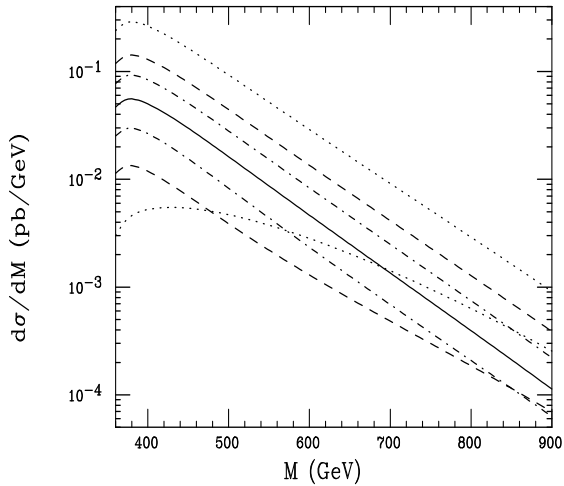,height=8.0cm,width=8.5cm,angle=-90}}
\vspace*{-0.6cm}
\caption{Top pair invariant mass distribution for top quark pairs produced at 
the Tevatron. The solid curve is the SM prediction and the upper(lower) 
dotted, dashed, and dash-dotted curves correspond to 
$\kappa=1,~0.5,~0.25(-1,~-0.5,~-0.25)$, respectively.}
\label{fig2}
\end{figure}
\vspace*{0.1mm}

We note that if $\tilde \kappa$ were also non-zero, the two parameters would be 
difficult to disentangle using measurements that do not probe for 
$CP$-violation since $\tilde \kappa$ also modifies the production cross 
section. Fig.2 from the analysis of K. Cheung~{\cite {old}} shows how the 
total cross section simultaneously varies with both of these parameters. 
Note that the presence of $\tilde \kappa$ can only lead to an increase in the 
cross section relative to the SM prediction and that the cross section 
is symmetric under $\tilde \kappa \to -\tilde \kappa$; this results from the 
fact that this parameter appears only in even powers in the various 
cross section expressions since its contribution cannot interfere with 
the pure SM piece due to $CP$. 

\vspace*{-0.5cm}
\nn
\begin{figure}[htbp]
\centerline{
\psfig{figure=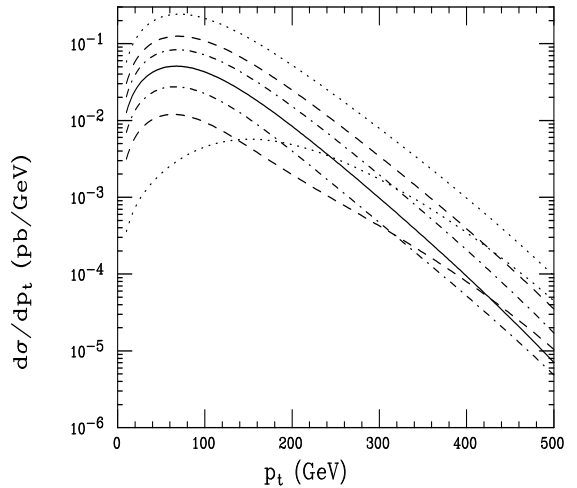,height=8.0cm,width=8.5cm,angle=-90}}
\vspace*{-0.6cm}
\caption{$p_t$ distributions of the top quark 
for the same $\kappa$ cases as shown in the previous figure.}
\label{fig3}
\end{figure}
\vspace*{0.1mm}

In an attempt to further probe $\kappa \neq 0$ we must explore the various 
kinematical distributions associated with top production. To this end we 
examine the top pair invariant mass distribution as well as the top quark $p_t$ 
distribution which are shown in Figs.3 and 4, respectively. When initially 
examining either of these figures one may at first believe that both of these 
distributions show a substantial sensitivity to $\kappa \neq 0$. However, 
a longer second 
look reveals that the sets of curves are mostly parallel especially if the 
magnitude of $\kappa$ is not too large. 

\vspace*{-0.5cm}
\nn
\begin{figure}[htbp]
\centerline{
\psfig{figure=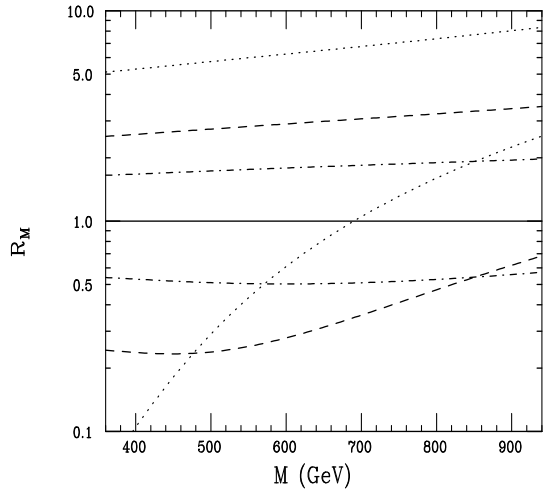,height=8.0cm,width=8.5cm,angle=-90}}
\vspace*{-0.6cm}
\caption{Ratios of the top pair invariant mass distributions for the various 
$\kappa$ values shown in Fig.3 to those of the SM.}
\label{fig4}
\end{figure}
\vspace*{0.1mm}

To see what this means 
let us form the cross section ratios $R_M$ and $R_{p_t}$ by dividing out the 
$\kappa$-dependent predictions by those of the SM; the results of this 
exercise are displayed in Figs.5 and 6. These figures show, for values of 
$\kappa$ in the range $|\kappa|\leq 0.25$ (and even for a somewhat larger 
interval), that the ratio of cross sections is at most slowly varying and 
is essentially constant, \ie, is 
independent of $M$ and/or $p_t$. Numerically, one finds that this 
constant ratio is just the ratio of the 
{\it total cross sections} observed in Fig.1. Thus we see that for relatively 
small, but phenomenologically interesting, values of $\kappa$ we learn 
no new information in this case from the $p_t$ or $M$ distributions except to 
verify the result of the total cross section measurements. 

\vspace*{-0.5cm}
\nn
\begin{figure}[htbp]
\centerline{
\psfig{figure=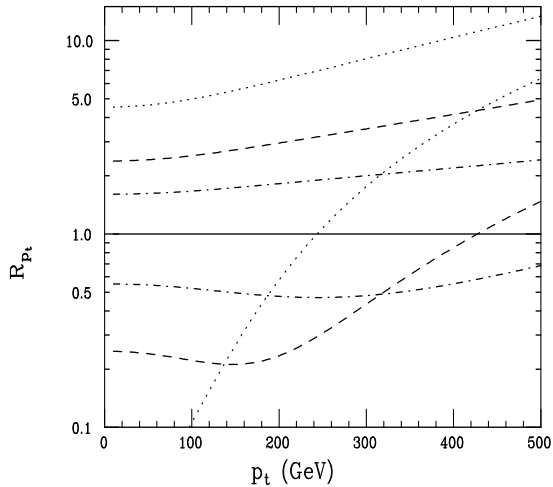,height=8.0cm,width=8.5cm,angle=-90}}
\vspace*{-0.6cm}
\caption{Ratios of the top quark $p_t$ distributions for the various 
$\kappa$ values shown in Fig.4 to those of the SM.}
\label{fig5}
\end{figure}
\vspace*{0.1mm}

What about other kinematic variables? Figs.7 and 8 display the top rapidity and 
angular distributions for the same set of $\kappa$ values as considered above. 
Unfortunately, it 
is again fairly obvious that for the relevant range of $\kappa$ the shapes of 
both of these distributions provide us with little or no extra sensitivity 
to $\kappa$ and again merely confirm the results from the total cross section. 

\vspace*{-0.5cm}
\nn
\begin{figure}[htbp]
\centerline{
\psfig{figure=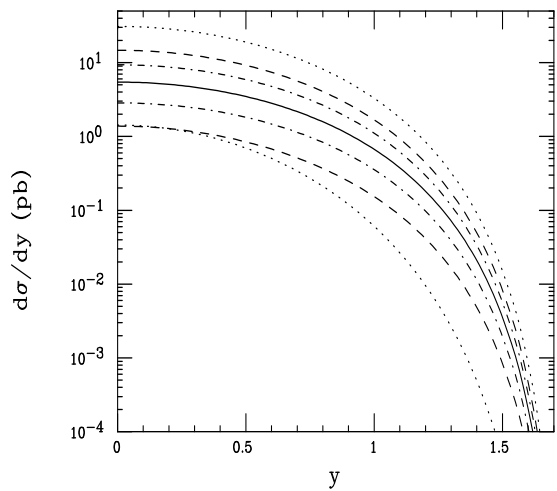,height=8.0cm,width=8.5cm,angle=-90}}
\vspace*{-0.6cm}
\caption{Top quark pair rapidity distributions for the same cases as shown 
in Fig.3.}
\label{fig6}
\end{figure}
\vspace*{0.1mm}

The above analysis shows that conventional top quark kinematical distributions 
are no more sensitive to $\kappa$ than is the total cross section itself. To 
try to bridge this problem, Cheung~{\cite {old}} examined the rate for 
single jet production, with fixed minimum 
$p_t$, in association with a top quark pair. 
This required the complete calculation of the $q\bar q,~gg\to t\bar tg$ and 
$qg\to t\bar tq$ subprocesses including the full $\kappa$ dependence. Cheung 
then compared the event rates for final states containing an additional jet 
with a fixed minimum $p_t$ as the value of $\kappa$ (and $\tilde \kappa$) was 
allowed to vary. For a minimum $p_t>20-25$ GeV the sensitivity of this 
reaction was found to be comparable to that obtained from the total cross 
section, although with completely different systematic errors. For 
lower minimum values of $p_t$ the sensitivity was reduced, whereas for 
higher minimum values of $p_t$ there is a loss in statistical power. It thus 
remains 
true that the Tevatron's sensitivity to $\kappa$ is essentially given by the 
bounds from the total cross section.

\vspace*{-0.5cm}
\nn
\begin{figure}[htbp]
\centerline{
\psfig{figure=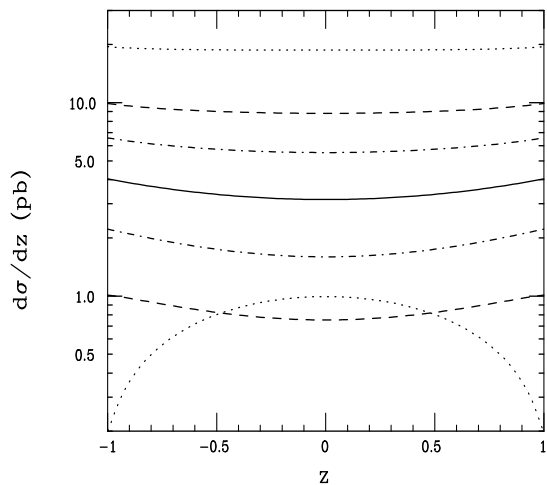,height=8.0cm,width=8.5cm,angle=-90}}
\vspace*{-0.6cm}
\caption{$z=cos \theta^*$ distribution for top-pair production as
in Figure 3, except that the upper(lower) dash-dotted, dashed,
and dotted curves correspond to $\kappa=0.25,~0.5,~1$$(-0.25,$~$~-0.5,~-
1)$,respectively.}
\label{fig7}
\end{figure}
\vspace*{0.1mm}

Before concluding this section we would like to briefly mention how the 
determination of $\kappa$ differs at the LHC where there is no shortage of 
statistics in top pair production. In this case it has been shown that the 
total cross section is somewhat less sensitive to $\kappa$ than at the 
Tevatron but the long lever arms in $M$ and, in particular, $p_t$ lead to 
substantially increased sensitivity. A preliminary study done during 
Snowmass '96~{\cite {old}} found that the LHC, through performing fits to 
these two distributions, was sensitive to values of $|\kappa|$ as small as 
0.04, which is about a factor of three better than the Tevatron.

\section{Low Scale Quantum Gravity}

Arkani-Hamed, Dimopoulos and Dvali(ADD)~{\cite {nima}} have recently 
proposed a radical solution to the hierarchy problem. ADD 
hypothesize the existence of $n$ additional large spatial dimensions in 
which gravity can live, called `the bulk', whereas all of the fields of the 
Standard Model are constrained to lie on `a wall', which is our 
conventional 4-dimensional world. Gravity only appears to be weak in our 
ordinary 4-dimensional space-time since we merely observe it's action on the 
wall. It has recently been shown~{\cite {nima}} that a 
scenario of this type may emerge in 
string models where the effective Planck scale in the bulk is identified 
with the string scale. In such a theory the hierarchy can be 
removed by postulating that the string or effective Planck scale in 
the bulk, $M_s$, is not far above the weak scale, \eg, a few TeV. Gauss' Law 
then provides a link between the values of $M_s$, the conventional 
Planck scale $M_{pl}$, and the size of the compactified extra dimensions, $R$, 
\begin{equation}
M_{pl}^2 \sim R^nM_s^{n+2}\,,
\end{equation}
where the constant of proportionality depends not only on the value of $n$ 
but upon the geometry of the compactified dimensions. Interestingly, if $M_s$ 
is near a TeV then $R\sim 10^{30/n-19}$ meters; for separations between two 
masses less than $R$ the gravitational 
force law becomes $1/r^{2+n}$. For $n=1$, 
$R\sim 10^{11}$ meters and is thus obviously excluded, but, for $n=2$ one 
obtains $R \sim 1$~mm, which is at the edge of the sensitivity for existing 
experiments. For $2<n \leq 7$, where 7 is the maximum value of $n$ 
being suggested by M-theory, the value of $R$ is further reduced and thus we 
may conclude that the range $2\leq n \leq 7$ is of phenomenological interest. 

\vspace*{-0.5cm}
\nn
\begin{figure}[htbp]
\centerline{
\psfig{figure=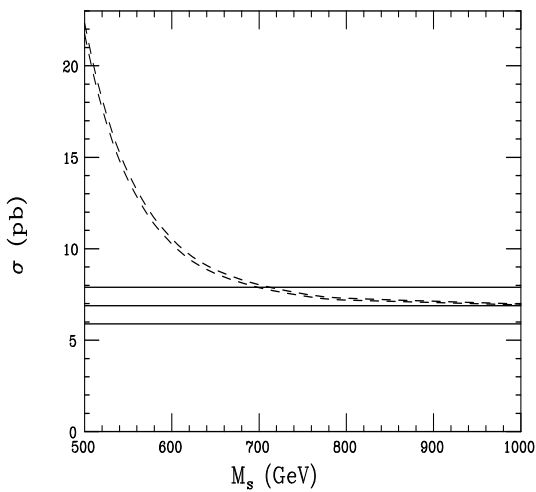,height=8.0cm,width=8.5cm,angle=-90}}
\vspace*{-0.6cm}
\caption{Top pair production cross section at Run II of the Tevatron as a 
function of the string scale $M_s$. The solid band represents an approximate 
$15\%$ error on the cross section determination. Although the results for 
both $\lambda=\pm 1$ are shown they are visually difficult to separate.}
\label{fig8}
\end{figure}
\vspace*{0.1mm}

The phenomenology of the ADD model has now begun to be addressed in a series 
of recent papers~{\cite {pheno}}. 
The Feynman rules can be obtained by considering a linearized theory of gravity 
in the bulk, decomposing it into the more familiar 4-dimensional states and 
recalling the existence of Kaluza-Klein towers for each of the conventionally 
massless fields. The entire set of fields in the K-K tower couples in an 
identical fashion to the particles of the SM. By considering the forms of the 
$4+n$  
symmetric conserved stress-energy tensor for the various SM fields and by 
remembering that such fields live only on the wall one may derive all of the 
necessary couplings. An important result of 
these considerations is that only the massive spin-2 K-K towers (which couple 
to the 4-dimensional stress-energy tensor, $T^{\mu\nu}$) and spin-0 K-K 
towers (which couple proportional to the trace of $T^{\mu\nu}$) are of 
phenomenological relevance as all the spin-1 fields can be shown to decouple 
from the particles of the SM. 
For processes that involve massless fermions at one vertex and massless 
gauge fields, as will be the case below, the contributions of the spin-0 
fields can also be ignored.

\vspace*{-0.5cm}
\nn
\begin{figure}[htbp]
\centerline{
\psfig{figure=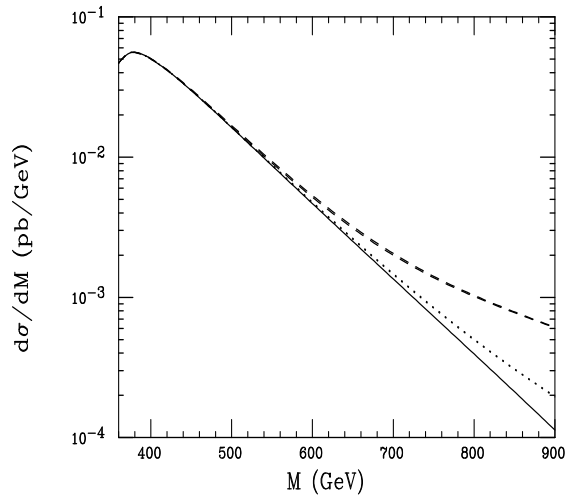,height=8.0cm,width=8.5cm,angle=-90}}
\vspace*{-0.6cm}
\caption{Top pair invariant mass distribution at Run II of the Tevatron 
assuming the SM(solid) or $M_s=800(1000)$ GeV as the dashed(dotted) curve.}
\label{fig9}
\end{figure}
\vspace*{0.1mm}

Given the Feynman rules as developed in {\cite {pheno}} it appears that the 
ADD scenario has two basic classes of collider tests. In the first class, 
a K-K tower of gravitons can be emitted during a decay or scattering process 
leading to a final state with missing energy. The rate for such processes is 
strongly dependent on the number of extra dimensions. 
In the second class, which we consider here, 
the exchange of a K-K graviton tower between SM fields can lead to almost 
$n$-independent modifications to conventional cross sections and 
distributions or can possibly lead to new interactions. The exchange of 
the graviton K-K tower leads to an effective color-singlet contact interaction 
operator of dimension-eight with a scale set by the parameter $M_s$ for 
both the $q\bar q \to t\bar t$ and $gg\to t\bar t$ processes. The single 
overall 
order one coefficient, $\lambda$, is unknown but its value is conventionally 
set to $\pm 1$. Given these two operators the relevant cross sections and 
distributions can be calculated directly.

For the process $q\bar q \to t\bar t$, we obtain 
\begin{eqnarray}
{d\sigma_{q\bar q}\over {d\hat t}}&=&{d\sigma^{SM}_{q\bar q}\over {d\hat t}}
+{\pi(\lambda K\hat s)^2\over 64 M_s^8}\left[1-3\beta^2 z^2
\right. \nonumber \\
&+&\left. 4\beta^4 z^4-(1-\beta^2)(1-4\beta^2 z^2)\right]\,,
\end{eqnarray}
where $\beta$, $M_s$ and $z$ have been defined above, which apart from a color 
factor was first obtained by Hewett~{\cite {pheno}} 
and which agrees with the expression given by Mathews, Raychaudhuri and 
Sridhar~{\cite {pheno}}. For $K=1({2\over \pi})$ we obtain the operator 
normalization 
employed by Guidice, Rattazzi and Wells(Hewett)~{\cite {pheno}}; for 
numerical analyses we will assume $K={2\over \pi}$ to make comparisons with 
other results obtained previously. Note that there is no 
interference between the SM contribution, which is pure color octet due to 
single gluon exchange, and that 
from the color-singlet graviton tower exchange; hence the sign of 
$\lambda$ is irrelevant in this case. Note that this term can only lead to an 
increase in the top production cross section. 

Similarly, for $gg\to t\bar t$, we find 
\begin{eqnarray}
{d\sigma_{gg}\over {d\hat t}}&=&{d\sigma^{SM}_{gg}\over {d\hat t}}
-{3\pi\over 64 \hat s^2}\left[{4(\lambda K)^2\over M_s^8}
\right. \nonumber \\
&-&\left. {8\alpha_s \lambda K\over 3M_s^4(m_t^2-\hat t)(m_t^2-\hat u)}\right]
\left[6m_t^8-4m_t^6(\hat t+\hat u)\right. \nonumber \\ 
&+&\left. 4m_t^2 \hat t \hat u(\hat t+\hat u)-\hat t\hat u(\hat t^2+\hat u^2)
+m_t^4(\hat t^2+\hat u^2\right. \nonumber \\
&-&\left. 6\hat t \hat u)\right]\,,
\end{eqnarray}
with $\hat t,\hat u={-1\over 2}\hat s(1\mp \beta z)+m_t^2$ and 
which agrees with the result obtained by 
Mathews, Raychaudhuri and Sridhar~{\cite {pheno}}. 
In this expression we see that an interference with the SM 
contribution does occur so that the sign of $\lambda$ is now relevant. This 
interference originates from the $t$- and $u$-channel SM graphs which have 
color singlet components. Numerically, the sign of $\lambda$ will remain 
relatively unimportant since the $gg\to t\bar t$ subprocess is suppressed at 
the Tevatron in comparison to $q\bar q \to t\bar t$. 

\vspace*{-0.5cm}
\nn
\begin{figure}[htbp]
\centerline{
\psfig{figure=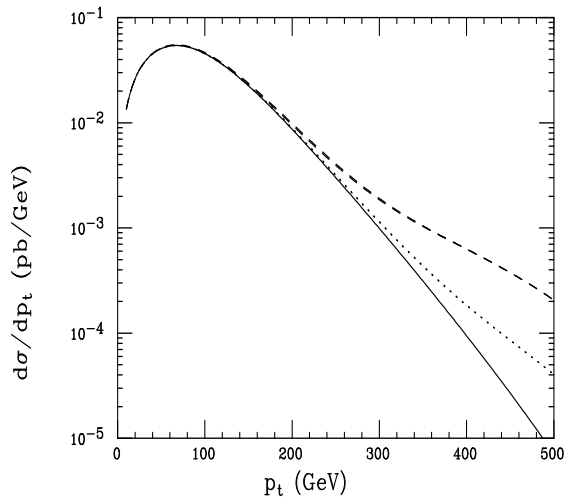,height=8.0cm,width=8.5cm,angle=-90}}
\vspace*{-0.6cm}
\caption{Same as the previous figure but now for the top $p_t$ distribution.}
\label{fig10}
\end{figure}
\vspace*{0.1mm}

In Fig.9 we show the top pair cross section for Run II as a function of $M_s$ 
for $\lambda=\pm 1$; these two curves are hardly separable with the difference 
due entirely to the SM-graviton tower interference term in the 
$gg\to t\bar t$ subprocess. Here we see that for $M_s$ less than about 700 
GeV the cross section becomes substantially larger than the SM prediction. 
Assuming that the SM value for the cross section was obtained experimentally 
would imply a value of $M_s$ 
greater than this value. This is consistent with the corresponding results of 
the Run I analysis by Mathews, Raychaudhuri and Sridhar~{\cite {pheno}}. 
To set the scale for this lower bound on $M_s$, we note that 
Hewett has shown~{\cite {pheno}} that the Tevatron at Run II, through the 
Drell-Yan process, should be sensitive to $M_s$ up to $\simeq 1.1$ TeV with a 
comparable sensitivity anticipated from LEP II. In addition, 
Rizzo~{\cite {pheno}} has shown that HERA may also eventually reach a similar 
level of sensitivity of $\simeq 1.1$ TeV. Evidently, we must go beyond the 
total cross section measurement~{\cite {jlhtgr} if we want to improve the 
$M_s$ reach using top production.

\vspace*{-0.5cm}
\nn
\begin{figure}[htbp]
\centerline{
\psfig{figure=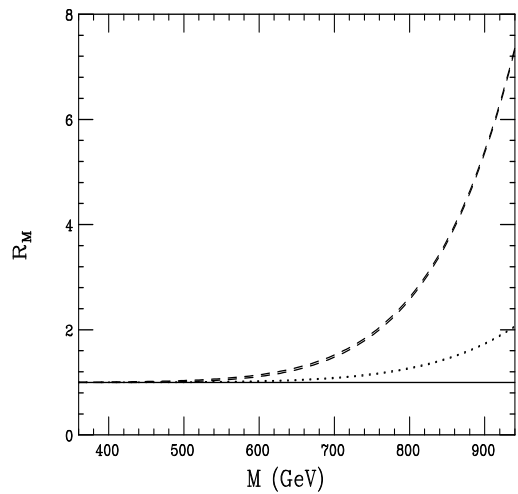,height=8.0cm,width=8.5cm,angle=-90}}
\vspace*{-0.6cm}
\caption{Ratios of the top pair invariant mass distributions for the two 
values of $M_s$ shown in Fig.10 to those of the SM.}
\label{fig11}
\end{figure}
\vspace*{0.1mm}

Figs.10 and 11 show the $t\bar t$ mass distribution and top quark $p_t$ 
distribution, respectively, for the SM and when $M_s=800$ or 1000 GeV. (The 
two sets of curves corresponding to $\lambda=\pm 1$ are shown but are not 
visually separable.) In the case of the top pair mass distribution no 
deviation from the SM is observed below invariant masses of $\simeq 600$ GeV 
where the three curves start to diverge. A similar situation is seen in the 
$p_t$ distribution below 200 GeV. To clarify the situation, we again form the 
ratios of predictions $R_M$ and $R_{p_t}$ shown in Figs.12 and 13. Both these 
ratios take off once the $M_s$ scale begins to be probed. (One may worry, based 
on unitarity arguments, that it is not valid to examine top pair invariant 
masses larger than the scale $M_s$. Guidice, Rattazzi and 
Wells~{\cite {pheno}} have shown that unitarity remains satisfied for values 
of $M \leq 1.6M_s$, using our normalization convention, so that all our plots 
are cut off before this point is reached.) 

\vspace*{-0.5cm}
\nn
\begin{figure}[htbp]
\centerline{
\psfig{figure=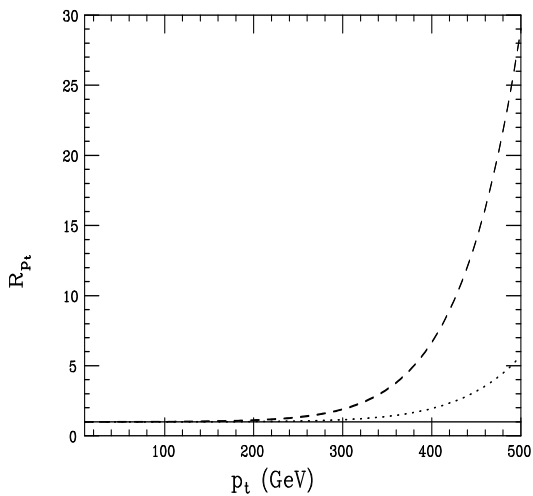,height=8.0cm,width=8.5cm,angle=-90}}
\vspace*{-0.6cm}
\caption{Ratios of the top quark $p_t$ distributions for the two 
values of $M_s$ shown in Fig.11 to those of the SM.}
\label{fig12}
\end{figure}
\vspace*{0.1mm}

To get an idea of the sensitivity of these distributions we have performed a 
simple Monte Carlo study. Assuming that the data reproduces the SM 
expectation and that the data below a certain cutoff can itself 
be used to normalize the distribution, then a
$95\%$ CL lower bound on $M_s$ can be obtained for a given fixed 
integrated luminosity. In the case of the mass distribution, we assume that 
the SM is applicable below $M=500$ GeV, as is reasonable from the figures, 
and fit Monte Carlo generated data to the $M_s$-dependent distribution for 
values of $M$ in excess of 600 GeV. Following this approach, for a 
integrated luminosity of 2(20) $fb^{-1}$ we obtain a $95\%$ CL lower limit on 
$M_s$ of 1.05(1.22) TeV with the sign of $\lambda$ dependence being less than 
$1\%$. Similarly, we assume that the $p_t$ distribution below 200 GeV is 
controlled by the SM and perform a $M_s$-dependent fit for $p_t>300$ GeV. For 
the luminosities above we obtain the corresponding $95\%$ CL lower bounds of 
$M_s>1.02(1.18)$ TeV, which are comparable to the previous results. Both sets 
of constraints are seen to be 
very similar to what is obtainable from Drell-Yan, LEP II and HERA data.

\vspace*{-0.5cm}
\nn
\begin{figure}[htbp]
\centerline{
\psfig{figure=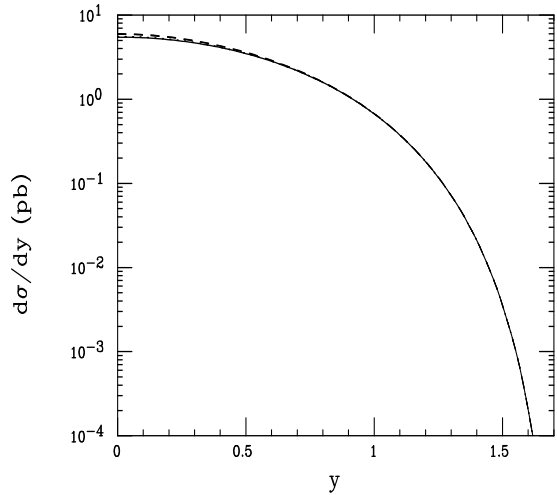,height=8.0cm,width=8.5cm,angle=-90}}
\vspace*{-0.6cm}
\caption{Top quark rapidity distribution at the Run II Tevatron for the 
SM(solid) and when the string scale $M_s$ is set to 800(1000) GeV 
corresponding to the dashed(dotted) curve.}
\label{fig13}
\end{figure}
\vspace*{0.1mm}

What about the rapidity and angular distributions? These are presented in 
Figs.14 
and 15. As can be seen, the rapidity distribution tells us essentially nothing 
since it is almost independent of $M_s$ for these values. Similarly, the 
angular distribution shown in Fig.15 displays only a very modest $M_s$ 
dependence. We conclude that little additional information on $M_s$ can be 
obtained from these distributions. 

\vspace*{-0.5cm}
\nn
\begin{figure}[htbp]
\centerline{
\psfig{figure=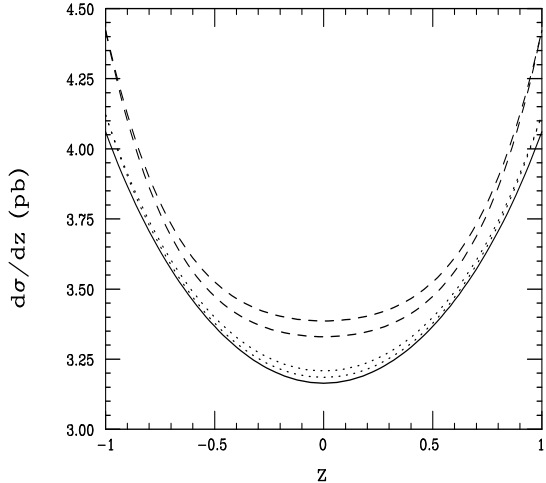,height=8.0cm,width=8.5cm,angle=-90}}
\vspace*{-0.6cm}
\caption{Top quark angular distribution at the Run II Tevatron for the 
SM(solid) and when the string scale $M_s$ is set to 800(1000) GeV 
corresponding to the dashed(dotted) curves. Note that the two cases 
$\lambda=\pm 1$ are separable in this plot.}
\label{fig14}
\end{figure}
\vspace*{0.1mm}

Before concluding this section we would again like to briefly mention the 
situation at the LHC~{\cite {jlhtgr}}. As one might expect 
the overall sensitivity to $M_s$ 
is here substantially increased and the sign of $\lambda$ becomes more 
important. The total cross section itself is found to be less sensitive to 
$M_s$; a $15\%$ measurement can probe values of $M_s$ only 
below $\simeq 1.8$ 
TeV. On the otherhand, fits to the $p_t$ and $M$ distributions similar to 
those discussed above yield {\it discovery} reaches as high as 5.7-6.3 TeV 
assuming an integrated luminosity of 100 $fb^{-1}$. This reach is comparable 
to that obtainable using the Drell-Yan process as shown by 
Hewett~{\cite {pheno}} and that which can be achieved at a 
1 TeV $e^+e^-$ linear collider.

\section{Summary and Conclusion}

The production properties of top quark pairs at the Tevatron can be uniquely 
sensitive to new physics beyond the SM. Here we have explored two types of new 
physics that appear as non-renormalizable, higher dimensional operators. 

In the first example, a dimension-six $P$ and $C$ conserving modification of 
the $t\bar tg$ vertex was shown to lead to an effective anomalous 
chromomagnetic moment for the top quark. At Tevatron energies this new 
interaction leads to essentially only one effect, \ie, the modification of 
the top pair production cross section. In this case all distributions, to first 
approximation, were shown to simply scale up or down by the same amount. 
Assuming the SM prediction for the cross section is obtained at Run II with 2 
$fb^{-1}$ of integrated luminosity we obtain the approximate bound 
$-0.10 \leq \kappa \leq 0.12$. 

In the case of low scale quantum gravity, the exchange of a Kaluza-Klein 
tower of gravitons leads to a pair of dimension-eight, color singlet operators 
that contribute to $t\bar t$ production with the effective Planck mass setting 
their scale. Unlike the case of the top quark anomalous 
chromomagnetic moment these operators induce distinctive modifications in both 
the total cross section as well as the top pair invariant mass and top 
transverse momentum distributions. By looking for 
deviations in these distributions at Run II we obtained sensitivities to the 
string scale of order 1 TeV. This can be increased by more than a factor of 
five in the case of the LHC.

\section*{Acknowledgements}
The author would like to thank Chris Carone for the opportunity to talk at 
this meeting and Fermilab for its hospitality. He would also like to thank 
J.L. Hewett, N. Arkani-Hamed, J. Wells, T. Han, J. Lykken, A. Kagan and 
D. Atwood for discussions related to this work.

\section*{References}
%
\def\MPL #1 #2 #3 {Mod. Phys. Lett. {\bf#1},\ #2 (#3)}
\def\NPB #1 #2 #3 {Nucl. Phys. {\bf#1},\ #2 (#3)}
\def\PLB #1 #2 #3 {Phys. Lett. {\bf#1},\ #2 (#3)}
\def\PR #1 #2 #3 {Phys. Rep. {\bf#1},\ #2 (#3)}
\def\PRD #1 #2 #3 {Phys. Rev. {\bf#1},\ #2 (#3)}
\def\PRL #1 #2 #3 {Phys. Rev. Lett. {\bf#1},\ #2 (#3)}
\def\RMP #1 #2 #3 {Rev. Mod. Phys. {\bf#1},\ #2 (#3)}
\def\EJPC #1 #2 #3 {E. Phys. J. {\bf#1},\ #2 (#3)}
\def\ZPC #1 #2 #3 {Z. Phys. {\bf#1},\ #2 (#3)}
\def\IJMP #1 #2 #3 {Int. J. Mod. Phys. {\bf#1},\ #2 (#3)}


\begin{thebibliography}{99}
%
\bibitem{nima}
N. Arkani-Hamed, S. Dimopoulos and G. Dvali, \PLB B429 263 1998 ~and 
hep-ph/9807344; I. Antoniadis, N. Arkani-Hamed, S. Dimopoulos and G. Dvali, 
\PLB B436 257 1998 ; N. Arkani-Hamed, S. Dimopoulos and J. March-Russell, 
hep-th/9809124; P.C. Argyres, S. Dimopoulos and J. March-Russell, 
hep-th/9808138; Z. Berezhiani and G. Dvali, hep-ph/9811378; N. Arkani-Hamed 
and S. Dimopoulos, hep-ph/9811353; Z. Kakushadze, hep-th/9811193 and 
hep-th/9812163; N. Arkani-Hamed \etal, hep-ph/9811448; 
G. Dvali and S.-H.H. Tye, hep-ph/9812483; 
G. Shiu and S.-H. H. Tye, \PRD D58 106007 1998 ;
Z. Kakushadze and S.-H. H. Tye, hep-th/9809147. See also, 
I. Antoniadis, \PLB B246 377 1990 ;
J. Lykken, \PRD D54 3693 1996 ;
E. Witten, \NPB B471 135 1996 ;
P. Horava and E. Witten, \NPB B460 506 1996 ~and \NPB B475 94 1996 ;
K.R. Diennes, E. Dudas and T. Gherghetta, \PLB B436 55 1998 ,~hep-ph/9803466, 
hep-ph/9806292 and hep-ph/9807522.
%
\bibitem{bing} 
K. Whisnant, J.M. Yang, B.-L. Young and X. Zhang, \PRD D56 467 1997 ;
J.M. Yang and B.-L. Young, \PRD D56 5907 1997 ;
K. Hikasa, K. Whisnant, J.M. Yang and B.-L. Young, hep-ph/9806401. 
%
\bibitem{est}
For some other estimates of the size of $\kappa$ in various extensions of the 
SM, see R. Martinez, J-A. Rodriguez and M. Vargas, hep-ph/9709478.
%
\bibitem{old}
D. Atwood, A. Kagan and T.G. Rizzo, \PRD D52 6264 1995 ;
K. Cheung, \PRD D53 3604 1996 ;
P. Haberl, O. Nachtmann and A. Wilch, \PRD D53 4875 1996 ;
T.G. Rizzo, \PRD D53 6218 1996 ;
T.G. Rizzo in {\it 1996 DPF/DPB Summer Study 
on New Directions for High Energy Physics}, Snowmass, CO, July 1996, eds. 
D.G. Cassel, L.Trindle Gennari and R.H. Siemann, hep-ph/9609311.
%
\bibitem{alex}
A.L. Kagan, \PRD D51 6196 1995 ;
M. Dine, A. Kagan and S. Samuel, \PLB B243 250 1990 .
%
\bibitem{veryold}
This is a generalization of the result first obtained by 
D. Silverman and G. Shaw, \PRD D27 1196 1983 ;
J. Reid, M. Samuel, K.A. Milton and T.G. Rizzo, \PRD D30 245 1984 .
%
\bibitem{cteq}
CTEQ Collaboration, \PRD D55 1280 1997 .
%
\bibitem{pheno}
G.F. Guidice, R. Rattazzi and J.D. Wells, hep-ph/9811291;
T. Han, J.D. Lykken and R.J. Zhang, hep-ph/9811350;
J.L. Hewett, hep-ph/9811356;
E.A. Mirabelli, M. Perlstein and M.E. Peskin, hep-ph/9811337;
P. Mathews, S. Raychaudhuri and K. Sridhar, hep-ph/9811501 and hep-ph/9812486; 
S. Nussinov and R.E. Shrock, hep-ph/9811323;
T.G. Rizzo, hep-ph/9901209.
%
\bibitem{jlhtgr}
For a complete analysis of the effects of graviton exchange on top production 
at both the Tevatron and LHC, see  J.L. Hewett and T.G. Rizzo, in preparation. 
%
\end{thebibliography}
\end{document}